\documentclass[%reprint,
preprint,
 aps,
prb,
]{revtex4-2}
\usepackage{graphicx}% Include figure files
\usepackage{dcolumn}% Align table columns on decimal point
\usepackage{bm}% bold math
\usepackage{xcolor}
\definecolor{red}{RGB}{200,0,0}
\newcommand{\ptb}{t-${\rm PtBi_2}$}
\usepackage{amsmath} % or simply amstext
\newcommand{\angstrom}{\text{\normalfont\AA}}
\begin{document}

\title{Topography of Fermi Arcs in t-PtBi$_2$ Using High Resolution Angle-resolved Photoemission Spectroscopy}

\author{Evan O'Leary$^{1,2}$, Zhuoqi Li$^{1,2}$, Lin-Lin Wang$^{1,2}$, Benjamin Schrunk$^{2}$, Andrew Eaton$^{1,2}$, Paul C. Canfield$^{1,2}$}
 \altaffiliation[ ]{canfield@ameslab.gov}
\author{Adam Kaminski$^{1,2}$}%
 \email{adamkam@ameslab.gov}
\affiliation{$^{1}$Iowa State University, Department of Physics and Astronomy, Ames, IA, 50014}

\affiliation{$^{2}$Ames Laboratory US Department of Energy, Ames, Iowa 50011, USA}

\date{\today}

\begin{abstract}
We use high resolution angle-resolved photoemission spectroscopy (ARPES) and density functional theory (DFT)  to investigate the electronic structure of trigonal phase ${\rm PtBi_2}$ (\ptb), a proposed Weyl semimetal that is expected to exhibit topological Fermi Arcs. Our ARPES data elucidates the topography of these objects and confirms their Fermi arc character. The arcs are formed by surface bands that have fairly flat bottom located very close to the chemical potential $\sim$ 6-8 meV, before they merge with bulk bands at higher binding energy. Comparison of the ARPES data with DFT calculations shows good agreement about their location and topography. Data acquired at low temperatures does not show any signatures of superconductivity down to 3 K in terms of expected changes in dispersion due to formation of Bogoliubov quasiparticles or superconducting gap in form of reliable shifts of peaks in energy distribution curves.

\clearpage

\end{abstract}

\maketitle

%\tableofcontents

\section{Introduction}
\noindent
    ${\rm PtBi_2}$ attracted significant attention because it was proposed to host topological  states. It is a polymorphic material, which in its cubic phase is a Dirac semimetal, that transforms to type-I Weyl state in its trigonal phase due to breaking of the inversion symmetry \cite{shipunov2020polymorphic,PhysRevB.99.161113}. More recently, there was a claim that the Fermi arcs present in the trigonal phase (\ptb) became superconducting below $\sim$ 10~K \cite{kuibarov2024evidence,PhysRevB.110.054504,veyrat2023berezinskii}. The superconducting surface state has been also reported by scanning tunneling microscopy with inhomogeneous superconducting gap ranging from 0-20 meV \cite{schimmel2024surface}. Signatures of the superconducting surface state were first reported by angle-resolved photoemission spectroscopy (ARPES) measurements \cite{kuibarov2024evidence}. 
 
Thus far, most of the focus on \ptb\ has been on the superconducting Fermi arcs, with less focus being placed on the actual topography of the bands giving rise to the arcs. Historically, the observation of Fermi arcs served as confirmation of a Weyl semimetal. Though it has been shown now that non-Weyl semimetals may exhibit Fermi arcs, such as in cuparates and rare-earth monopnictides  \cite{kaminski2015pairing,schrunk2022emergence,kushnirenko2022rare,kushnirenko2024unexpected,kushnirenko2023directional}. In the past, the investigation of such states was complicated by very close proximity of other trivial bands such in TaAs, TaP, NbAs, and NbP \cite{lv2015experimental,yang2015weyl,xu2016observation,xu2015discovery,souma2016direct}. The Fermi arcs in \ptb\ are well separated from other bands, so this material presents an ideal opportunity to study their topography. Previous studies of this material relied mostly on the DFT calculation and data showed only streaks of intensity at the expected locations of the arcs, which were not sufficient to study their dispersion nor demonstrate that they are indeed open contours in momentum space. We also demonstrate the absence of changes in the electronic properties that would indicate presence of features expected in the superconducting state. Our results point to the fact that superconductivity is not a universal feature in this material and further studies are necessary to identify the additional conditions necessary for its appearance if indeed present.

\section{Methods}
\noindent
%Crystal growth method description
Single crystals of the hP9/P-3, metastable phase of ${\rm PtBi_2}$ were grown in a manner similar to that used to grow metastable, triclinic, ${\rm RhBi_2}$ \cite{lee2021discovery}. Elemental Bi and Pt were combined in a ${\rm Bi_{80}Pt_{20}}$ ratio and placed in a fritted, alumina crucible set (sold as Canfield Crucible Set or CCS) \cite{canfield2016use,crucible}. The CCS was sealed into an amorphous silica tube and placed into a box furnace \cite{canfield2019new}. The ampoule was then heated to 1000 °C over 5 hours, held at 1000 °C for 5 hours, cooled to 700 °C over 4 hours and then slowly cooled to 440 °C over 146 hours. At 440 °C, above the reported 420 °C transformation temperature to the cP12/Pa-3 phase of ${\rm PtBi_2}$ [ASM phase diagram 979935 for Bi-Pt binary phase diagram], the ampoule was removed from the furnace and the excess, Bi-rich solution was decanted \cite{canfield2019new}. Once the ampoule cooled to room temperature it was opened and large, cleanly faceted plates (see inset to Fig.~\ref{fig:DFT} (b)) of hP9/P-3 ${\rm PtBi_2}$ were revealed. X-ray diffraction measurements taken along [001] direction, shown in Fig.~\ref{fig:DFT} (b), made on ground single crystals confirmed the correct, high temperature phase and further illustrated the stability of the phase at room temperature.

%ARPES description here
The ARPES measurements were performed by cleaving \ptb\ \textit{in situ} at base pressure lower than 5 $\times$ 10$^{-11}$ Torr. The 330 K and 360 K data were measured from the same sample which was cleaved at 330 K. The low temperature data were measured on samples that were cleaved at 4 K. A Scienta DA30 analyzer was used, in conjunction with a tunable, picosecond Ti:Sapphire laser and fourth harmonic generator \cite{jiang2014tunable}. A second fixed wavelength laser was used, with second harmonic generator, and Scienta DA30 analyzer. The laser photon energies were 6.7 eV and 7 eV for Ti:Sapphire and fixed wavelength lasers, respectively. The chemical potential was determined by measuring a Cu sample that was in electrical contact with the \ptb, and fitting a Fermi function to the data. A He discharge lamp and Scienta R8000 were used to obtain wider momentum space Fermi surface maps, at a photon energy of 21.2 eV. Energy resolution was set at 1 meV and 2 meV for laser and He discharge lamp, respectively. Peak positions were extracted by fitting Lorentzians to energy distribution curve (EDC) data.

%DTF calculation section 
Band structure of trigonal ${\rm PtBi_2}$ in space group P31m (157) with spin-orbit coupling (SOC) in density functional theory (DFT) have been calculated with PBE exchange-correlation functional, a plane-wave basis set and projected augmented wave method as implemented in VASP \cite{hohenberg1964inhomogeneous_DFT1,kohn1965self_DFT2,perdew1996generalized_DFT3,blochl1994projector_DFT4,kresse1996efficient_DFT5,kresse1996efficiency_DFT6}. The experimental lattice parameters of a = 6.5730 \angstrom\ and c = 6.1665 \angstrom\ have been used in the DFT calculations \cite{kaiser2014bi2pt}. A Monkhorst-Pack (8×8×8) k-point mesh with a Gaussian smearing of 0.05 eV including the $\Gamma$ point and a kinetic energy cutoff of 230.3 eV have been used \cite{monkhorst1976special_DFT8}. A tight-binding model based on maximally localized Wannier functions is constructed to reproduce closely the bulk band structure including SOC in the range of $E_F$ ± 1 eV with Pt \textit{sd} and Bi \textit{p} orbitals \cite{marzari1997maximally_DFT9,souza2001maximally_DFT10,marzari2012maximally_DFT11}. The surface spectral functions of the semi-infinite surface have been calculated with the surface Green’s function methods as implemented in WannierTools \cite{sancho1984quick_DFT14,sancho1985highly_DFT15,wu2018wanniertools_DFT16}. 

\section{Results and discussion}
\begin{figure}
\centering
\includegraphics[scale=0.50]{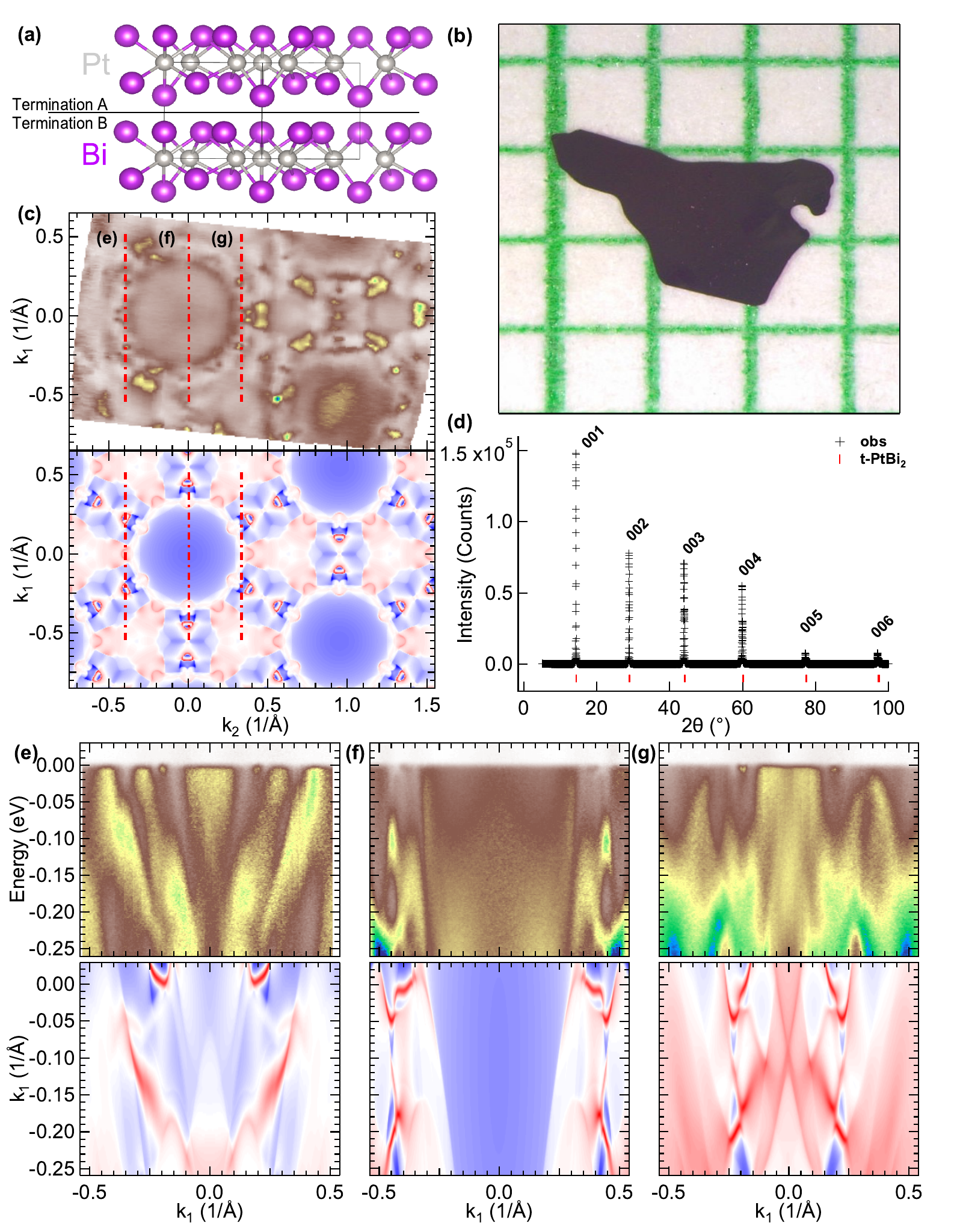}
    \caption{Density functional theory calculations and ARPES data from \ptb\ samples. (a) Crystal structure of \ptb, solid line indicates cleaving plane. (b) Photograph of \ptb\ sample on mm grid paper. (c) Fermi surface comparison for termination A between ARPES (top) and DFT (bottom), ARPES data is integrated within 10 meV of the Fermi level and taken at 7 K. (d) XRD intensity showing observed points as plus sign (+) and expected peak locations for \ptb\ as a vertical bar ($\mid$), taken along [001] direction. (e)-(g) ARPES (left) and DFT (right) energy dispersions along the lines indicated in panel (b).}
    \label{fig:DFT}
\end{figure}

\begin{figure}
\centering
\includegraphics[scale=0.52]{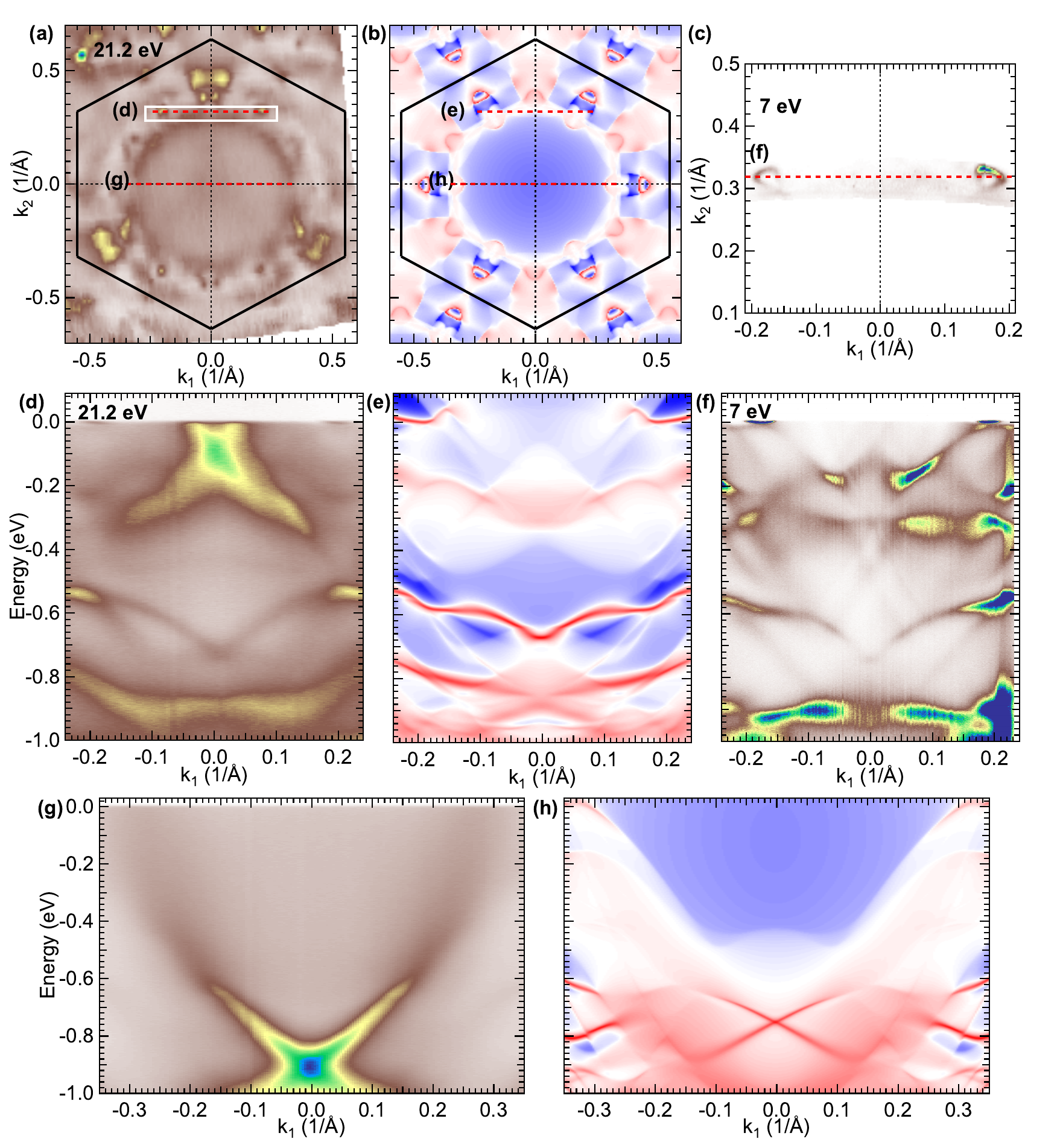}
    \caption{Comparison between data and calculations of Fermi surface maps and energy dispersions for termination A. (a) ARPES intensity integrated within 10 meV of the Fermi level, using a photon energy of 21.2 eV and taken at 6 K. (b) Fermi surface calculated by DFT. (c) ARPES intensity shown within 1 meV of the Fermi level measured using a photon energy of 7 eV, at 4 K. Data location is outlined by white rectangle in (a). (d)-(h) Energy dispersion along the lines shown in the above Fermi surface maps.}
    \label{fig:LaserHeDFT}
\end{figure}

\noindent
The trigonal phase of ${\rm PtBi_2}$ cleaves between two Bi layers, shown in Fig.~\ref{fig:DFT} (a), resulting in two different surface terminations. Termination A is composed of Bi atoms in a buckled formation shown above the line in Fig.~\ref{fig:DFT} (a), and termination B, has Bi atoms in a flat formation, shown below the line. Electronic structures of both terminations were studied previously by DFT and ARPES \cite{kuibarov2024evidence}, here we focus our attention on termination A, which the previous study found to have higher T$_c$ of $\sim$14~K. In order to select desired termination, samples were split in half and mounted with the opposing sides facing up.  We then compared measured Fermi surface with DFT calculations as shown in Fig.~\ref{fig:DFT} (c). The most apparent similarity is a large circular pocket centered around $\Gamma$. Closer inspection reveals smaller similarities, extending even to the second Brillouin zone. The trigonal nature of \ptb\ is more apparent in the ARPES data than the DFT calculations. Inspection of the ARPES Fermi surface map shows a clear three-fold symmetry, while DFT calculations appear to have a six-fold symmetry at first sight. Even with this difference, the ARPES data and DFT calculations are in good agreement. In Fig.~\ref{fig:DFT} (e) we can see in the ARPES data a Dirac like dispersion with a crossing around 0.12 eV below the Fermi level. The ARPES data shown in Fig.~\ref{fig:DFT} (f) has no apparent bands near the center of the zone, but along the edges shows some sharper features. In the final cut shown in Fig.~\ref{fig:DFT} (g) some broad bands are in the center of the cut, with some sharper bands around 0.25 1/\angstrom. The features discussed in the ARPES data are all present in the DFT calculations. In Fig.~\ref{fig:LaserHeDFT} (g) ARPES data shows a sharp band crossing around 0.9 eV, this cut matches the cut shown in Fig.~\ref{fig:DFT} (f), but shows a wider energy range. The DFT calculation shown in Fig.~\ref{fig:LaserHeDFT} (h) shows a matching band crossing to the one in the ARPES data. This sharp band crossing was absent from DFT calculations for termination B. From these comparisons, the ARPES data demonstrates good agreement with the DFT calculations. One final thing of note in Fig.~\ref{fig:DFT} (e)-(g) is the appearance of what we later show to be Fermi arcs. They can be seen in the ARPES data just below the Fermi level in all three cuts. From the comparison between the ARPES data and DFT calculations, we conclude the ARPES data corresponds to termination A.

To investigate the topography of the Fermi arcs we use high resolution data measured using fixed wavelength and tunable laser spectrometer. In Fig.~\ref{fig:LaserHeDFT} (c) we show a high resolution intensity map at the chemical potential measured using a photon energy of 7 eV. This map shows two Fermi sheet contours, both of which have a horseshoe like shape, although it is not possible to conclude whether or not they are open or closed (i.e. are Fermi arcs vs Fermi pockets). In order to establish the location of the map shown in Fig.~\ref{fig:LaserHeDFT} (c), we compared the ARPES data taken at 21.2 eV with DFT calculations. The location of the data in Fig.~\ref{fig:LaserHeDFT} (c) is shown by the white box in Fig.~\ref{fig:LaserHeDFT} (a). This was concluded by comparing energy dispersion cuts (d)-(f). In all three cuts, a band is centered around a binding energy of 0.65 eV and a second one is centered around a binding energy of 0.9 eV. The key features present in both data sets allow us to conclude that they are from the same termination (A) and location and, agree with DFT calculations. While there are some differences between the two data sets, this is to be expected as previous ARPES data on \ptb\ shows strong photon energy dependence \cite{kuibarov2024evidence}. In Fig.~\ref{fig:LaserHeDFT} (d) the arcs appear very faintly, but can be seen near the Fermi level. 

\begin{figure}
\centering
\includegraphics[scale=0.6]{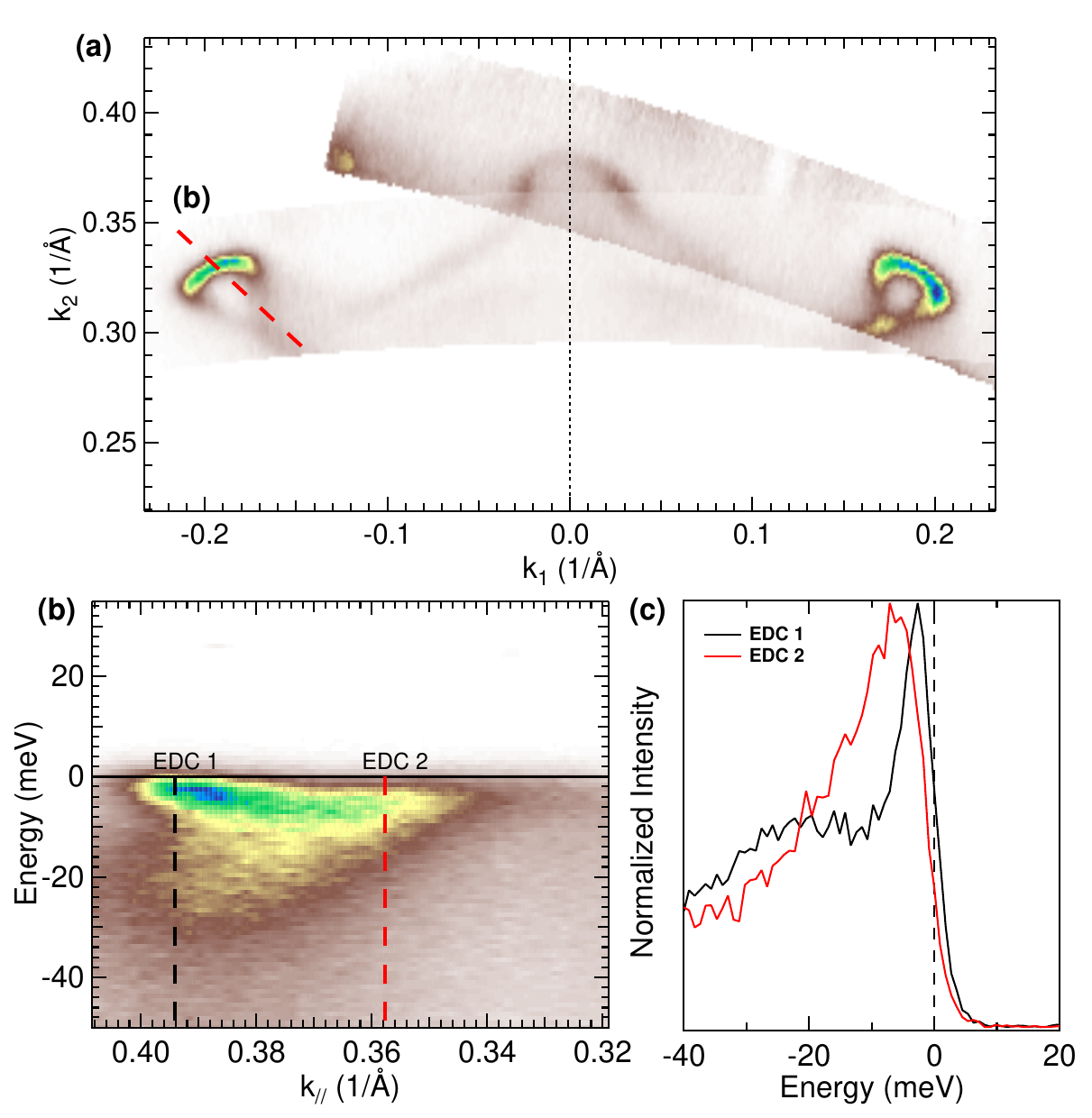}
    \caption{Detailed examination of band dispersion that forms Fermi arcs for termination A, taken at 4 K. (a) ARPES intensity shown within 1 meV of the Fermi level measured using a photon energy of 7 eV. Map shows two datasets obtained with two different sample rotations, superimposed on top of each other. (b) Band dispersion through the Fermi arc along the line shown in (a). (c) EDC along the dashed lines marked in (b).}
    \label{fig:Pocket2}
\end{figure}

The key question that needs to be addressed is whether or not the Fermi arcs predicted by calculations are indeed disconnected contours in momentum space or are they closed Fermi pockets in real data. We are referring to the objects shown explicitly in Fig.~\ref{fig:LaserHeDFT} (c). Previous ARPES data from NbAs demonstrated that that such objects may appear closed, but are actually open Fermi arcs upon careful examination \cite{xu2015discovery}. This may occur when the splitting of Weyl points is small and they are located close together in momentum space. It may also be due to the localization of the Fermi arcs near the Fermi level. In ARPES data, this localization may cause a convolution of the Fermi arc with the Fermi function which leads to the appearance of a band crossing, where there is none. We examine this situation for \ptb\ in more detail in Fig.~\ref{fig:Pocket2} (a), where we plot ARPES intensity map enclosing two of the suspected Fermi arcs. The band dispersion plot along the red dashed line cutting through the Fermi arc, is shown in Fig.~\ref{fig:Pocket2} (b). This plot demonstrates that the left side of the band clearly crosses the chemical potential near $k_{//}$ = 0.4 $(1/\angstrom)$, while the right side has no apparent crossing and the band loses intensity before crossing the chemical potential. The ultimate proof for absence of crossing on the right side is provided by examining the EDCs shown in Fig.~\ref{fig:Pocket2} (c). The black EDC obtained at the maximum of intensity at the chemical potential on the left side shows a peak located very close to zero energy, with the leading edge crossing it well above the half-way point. This is the expected shape and location of an EDC peak at the Fermi crossing. The red EDC, measured the same way on the right side the the band, shows a peak that is located at higher binding energy with the leading edge shifted by several meV's from the black curve. This demonstrates that the right side of the band does not cross the Fermi level, and instead it loses intensity by merging with the bulk band. This is as expected due to the properties of bands that form Fermi arcs. This suggests that the contours shown in Fig.~\ref{fig:Pocket2} (a) are indeed Fermi arcs, and provides experimental evidence that \ptb\ is indeed a Weyl semimetal.

\begin{figure}
\centering
\includegraphics[scale=0.6]{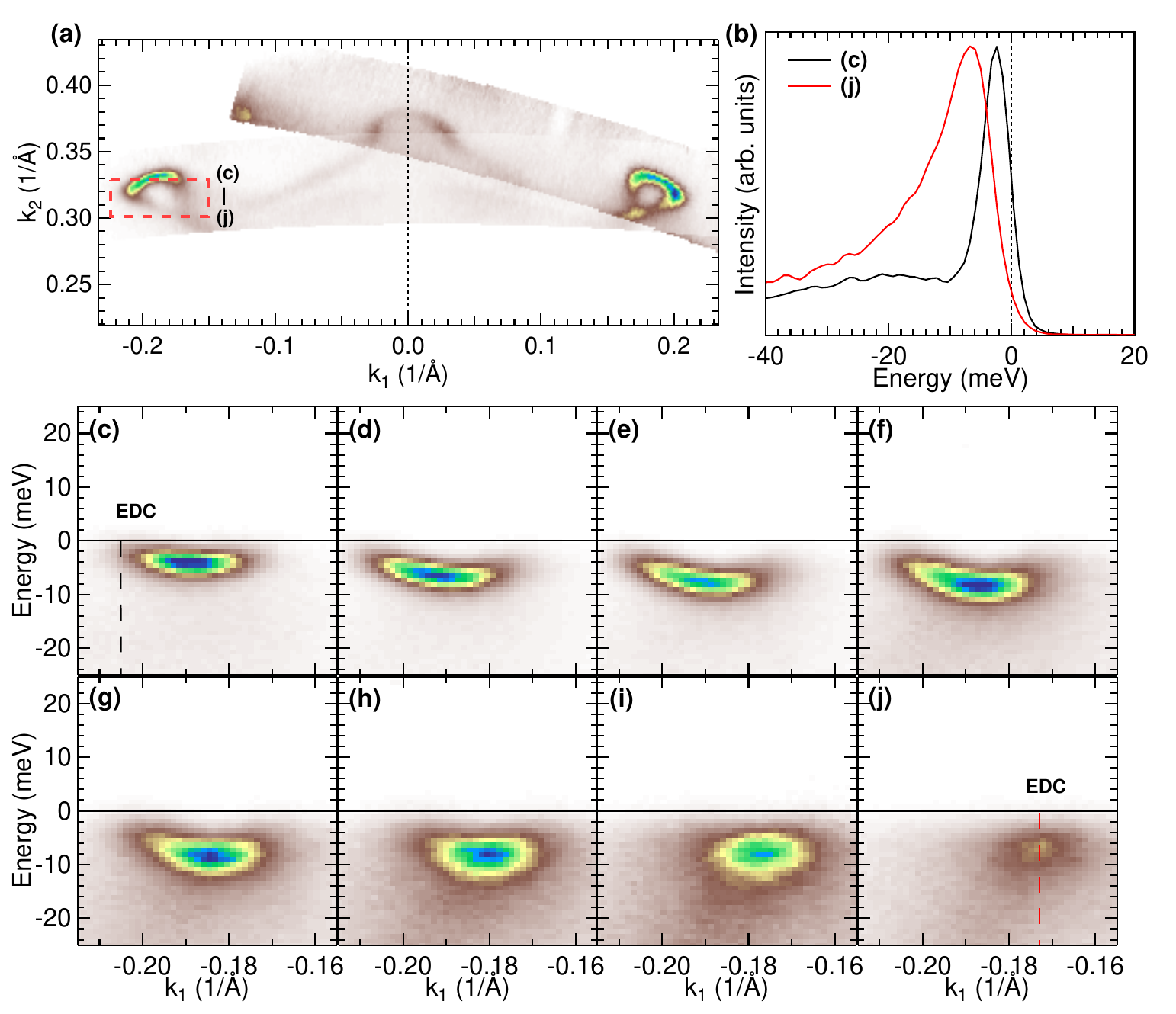}
    \caption{Analysis of shape and band dispersion giving rise to Fermi arcs for termination A, taken at 4 K. (a) ARPES intensity integrated within 1 meV of the Fermi level measured using photon energy of 7 eV. Map shows two datasets taken at different sample rotations, superimposed on top of each other. (b) Energy distribution curve along lines shown in (c) and (j). (c)-(j) ARPES intensity plots along horizontal cuts within the red rectangle marked in (a).}
    \label{fig:Pocket}
\end{figure}

Extensive mapping of the Fermi arcs is shown in Fig.~\ref{fig:Pocket}. The ARPES dispersion data, shown in Fig.~\ref{fig:Pocket} (c)-(j), are plotted along horizontal cuts through the red rectangle indicated on the Fermi map in Fig.~\ref{fig:Pocket} (a). At the top cut, Fig.~\ref{fig:Pocket} (c), the bottom of the band the gives rise the Fermi arc is at its lowest binding energy i.e. it is closest to the Fermi level and the band crosses the Fermi level on both sides. As we move further down the highlighted region, the bottom of this band shifts to slightly higher binding energy and at the same time the band no longer crosses the Fermi level on the right side, instead its intensity vanishes along the horizontal direction. Eventually both sides of the band lose intensity before reaching the Fermi level, as evident from data in panels (h-j). This behavior is also illustrated by EDCs shown in Fig.~\ref{fig:Pocket} (b). The black curve is measured at dashed line indicated in panel (c) and shows clear crossing of the band as the Fermi level is well above half intensity of the leading edge. In contrast, the EDC measured along the red dashed line in (j) has a peak located $\sim$ 7 meV below the Fermi energy, demonstrating the lack of Fermi crossing. This is further confirmation that these pockets are indeed Fermi arcs, that is, the intensity contour in panel (a) has a crossing on upper left side and does not have crossing on the lower right side. The energy dispersions in Fig.~\ref{fig:Pocket} (c)-(j) also show that the band giving rise to the Fermi arcs are very shallow, only $\sim$ 10 meV below the Fermi level. This energy scale can present challenges in observing potential energy gaps due to superconductivity. 

\begin{figure}
\centering
\includegraphics[scale=0.50]{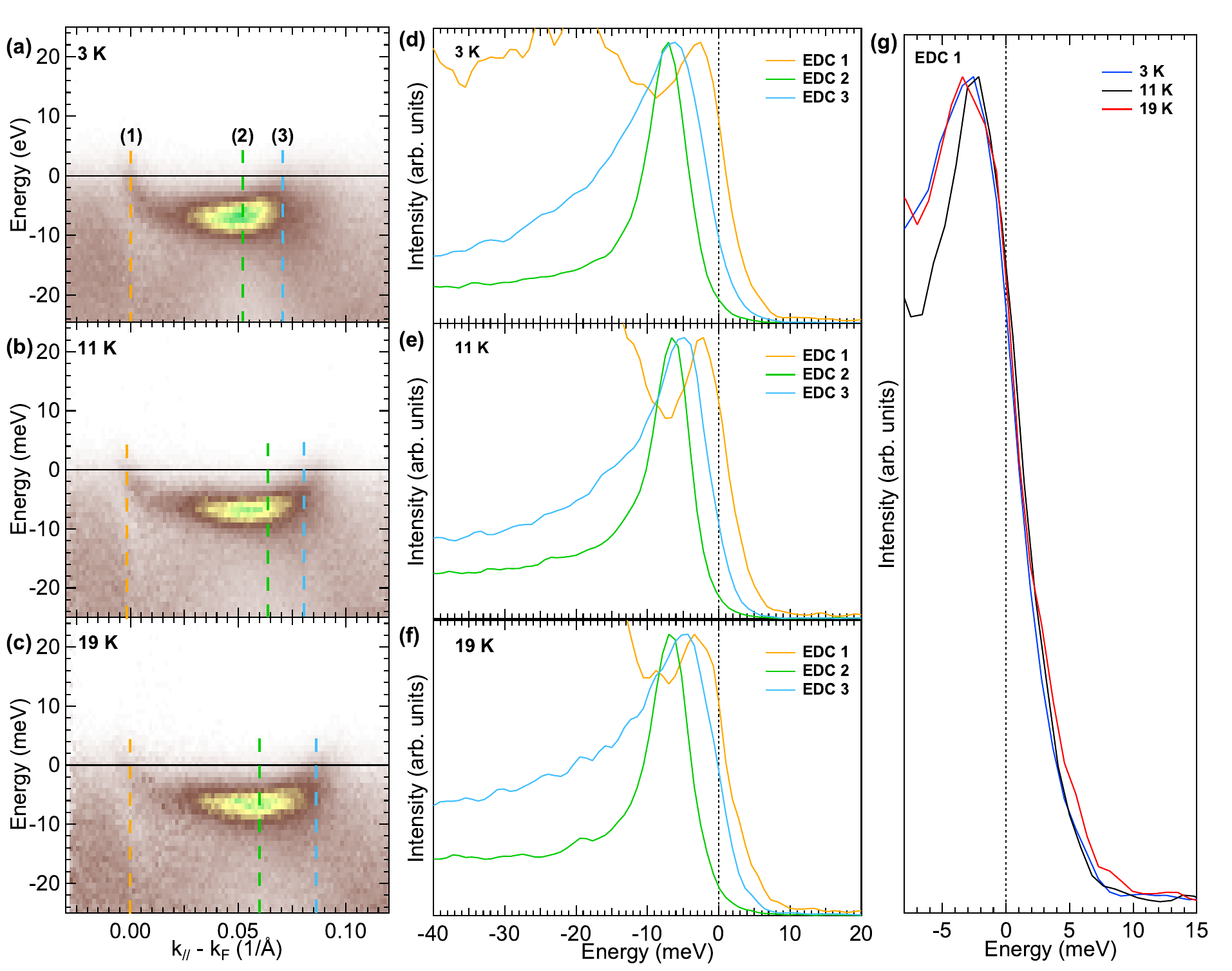}
    \caption{Temperature evolution of the Fermi Arcs on termination A. (a)-(c) Zoomed in ARPES intensity plot through Fermi arc.  (d)-(f) EDCs along dashed lines marked in (a)-(c) measured at 3~K, 11~K and 19~K. (g) Comparison of EDCs at the Fermi crossing measured at the three temperatures}
    \label{fig:Temp}
\end{figure}

The electronic structure of trigonal ${\rm PtBi_2}$ has many interesting spectral features including recent reports of observation of superconducting gap at the Fermi arcs \cite{kuibarov2024evidence,PhysRevB.110.054504}. The reported superconducting gap was present only in the surface states, and was not observed in the bulk bands. Theoretical understanding of such gap is still lacking. Some proposals suggest that inversion symmetry breaking, coexistence of spin singlet and triplet pairing may be responsible for the surface superconductivity reported experimentally \cite{PhysRevB.110.054504}. The superconducting transition temperature is reported to be different for termination A and termination B. For one termination $T_{cA}$ = 14 K and for the other termination $T_{cB}$ = 8 K \cite{kuibarov2024evidence}. We attempted to reproduce these results for termination A by performing detailed temperature dependent ARPES measurements of the Fermi arcs down to 3~K. We summarize our results in Fig.~\ref{fig:Temp}. Panels (a-c) show ARPES intensity maps measured at 3~K, 11~K and 19~K, respectively. We note that there are no observable changes in the data between different temperatures from above to below expected T$_c$. There is no characteristic back bending near Fermi crossing due to formation of Bogoliubov quasi-particles, nor any other observable changes that would indicate transition to the superconducting state and formation of a gap. We also note that the intensity is very high close to the band minimum, and becomes significantly weaker as the band crosses the chemical potential.

 In Fig.~\ref{fig:Temp} (d)-(f) we show EDCs plotted at momenta indicated by dashed lines in panels (a)-(c) that correspond to (1) the band crossing i.e. Fermi momentum, (2) the band local minimum, and (3) part of the band on the opposite side to the band contour from the arc where band loses intensity without crossing the Fermi momentum. These EDCs were normalized to the peak intensity for easier comparison of the location of the leading edge.  The band local minimum at this cut is located $\sim$7~meV below the chemical potential. The use of tunable laser in our measurements as well as previous reports poses a two fold challenge for the measurement of precise energy of the spectral features \cite{kuibarov2024evidence}. First, tunable Ti:sapphire lasers have inherent wavelength temperature instability and very small changes in wavelength result is meV scale shifts in photon energy and therefore determination of the chemical potential. Second, due to space charge effects caused by repulsion between photoelectrons within the bunch, and small changes of beam intensity common for Ti:Sapphire and fixed wavelength lasers, result in substantial shifts in photoelectron energies, and thus again in determination of the chemical potential \cite{graf2010vacuum}. Even with use of 3 ps pulses in our case, shifts of order of meV are common. To combat this we used frequent measurements of the chemical potential reference and tried to perform measurements within a short period of time. Even then, the data set measured at the lowest temperature was shifted by $\sim$1 meV  above the chemical potential - in a direction opposite to what is expected if the superconducting gap is present. We took these variations into account by using the EDC~2 as a reference. Since the local band minimum is below the chemical potential, it is not substantially affected by temperature changes within the very limited temperature range of these measurements and provides intrinsic reference for the chemical potential that is self compensating for any experimental shifts due to variation of laser wavelength or space charge effects. After such corrections, the EDC at the Fermi momentum remains unchanged in our experiment down to the lowest accessible temperature of 3~K, as shown in Fig.~\ref{fig:Temp} (g). Close inspection of EDCs on multiple samples lead us to conclude that all the data from our samples does not show any evidence of a superconducting gap. The band clearly crosses the chemical potential at each temperature shown, indicating absence of an energy gap. This was verified using several samples.

\section{Conclusions}
\noindent In summary, we presented high resolution ARPES data and DFT calculations for type-I Weyl semimetal \ptb\ exploring the properties of topologically protected surface states. The ARPES data agrees quite well with the DFT calculations, and allowed us to conclude that we were examining the buckled Bi surface A termination. Careful analysis of the ARPES data allowed us to demonstrate that the seemingly closed pockets of intensity are actually real Fermi arcs, and are disconnected contours in momentum space. This was evident from the fact that the band giving rise to these contours crossed the chemical potential only on one side and lost intensity on the opposite side, merging with the bulk band. These conclusions were verified also by examination of EDCs at several points of the contours. Finally we investigated the Fermi arcs for the presence of a superconducting gap. We found no evidence for presence of Bogoliubov quasiparticles, changes in band dispersion, energy shifts of EDC peaks or leading edge that would indicate the formation of a superconducting gap. Based on this we can conclude that Fermi arcs in our samples are not superconducting down to 3 K. 

\section{acknowledgments}
\noindent This work was supported by the U.S. Department of Energy, Office of Basic Energy Sciences, Division of Materials Science and Engineering, Ames National Laboratory which is operated for the U.S. Department of Energy by Iowa State University under Contract No. DE-AC02-07CH11358. First-principles calculations used resources of the National Energy Research Scientific Computing Center (NERSC), a DOE Office of Science User Facility.
\clearpage

\clearpage

\bibliography{apssamp}% Produces the bibliography via BibTeX.

\providecommand{\noopsort}[1]{}\providecommand{\singleletter}[1]{#1}%
\begin{thebibliography}{36}%
\makeatletter
\providecommand \@ifxundefined [1]{%
 \@ifx{#1\undefined}
}%
\providecommand \@ifnum [1]{%
 \ifnum #1\expandafter \@firstoftwo
 \else \expandafter \@secondoftwo
 \fi
}%
\providecommand \@ifx [1]{%
 \ifx #1\expandafter \@firstoftwo
 \else \expandafter \@secondoftwo
 \fi
}%
\providecommand \natexlab [1]{#1}%
\providecommand \enquote  [1]{``#1''}%
\providecommand \bibnamefont  [1]{#1}%
\providecommand \bibfnamefont [1]{#1}%
\providecommand \citenamefont [1]{#1}%
\providecommand \href@noop [0]{\@secondoftwo}%
\providecommand \href [0]{\begingroup \@sanitize@url \@href}%
\providecommand \@href[1]{\@@startlink{#1}\@@href}%
\providecommand \@@href[1]{\endgroup#1\@@endlink}%
\providecommand \@sanitize@url [0]{\catcode `\\12\catcode `\$12\catcode `\&12\catcode `\#12\catcode `\^12\catcode `\_12\catcode `\%12\relax}%
\providecommand \@@startlink[1]{}%
\providecommand \@@endlink[0]{}%
\providecommand \url  [0]{\begingroup\@sanitize@url \@url }%
\providecommand \@url [1]{\endgroup\@href {#1}{\urlprefix }}%
\providecommand \urlprefix  [0]{URL }%
\providecommand \Eprint [0]{\href }%
\providecommand \doibase [0]{https://doi.org/}%
\providecommand \selectlanguage [0]{\@gobble}%
\providecommand \bibinfo  [0]{\@secondoftwo}%
\providecommand \bibfield  [0]{\@secondoftwo}%
\providecommand \translation [1]{[#1]}%
\providecommand \BibitemOpen [0]{}%
\providecommand \bibitemStop [0]{}%
\providecommand \bibitemNoStop [0]{.\EOS\space}%
\providecommand \EOS [0]{\spacefactor3000\relax}%
\providecommand \BibitemShut  [1]{\csname bibitem#1\endcsname}%
\let\auto@bib@innerbib\@empty
%</preamble>
\bibitem [{\citenamefont {Shipunov}\ \emph {et~al.}(2020)\citenamefont {Shipunov}, \citenamefont {Kovalchuk}, \citenamefont {Piening}, \citenamefont {Labracherie}, \citenamefont {Veyrat}, \citenamefont {Wolf}, \citenamefont {Lubk}, \citenamefont {Subakti}, \citenamefont {Giraud}, \citenamefont {Dufouleur} \emph {et~al.}}]{shipunov2020polymorphic}%
  \BibitemOpen
  \bibfield  {author} {\bibinfo {author} {\bibfnamefont {G.}~\bibnamefont {Shipunov}}, \bibinfo {author} {\bibfnamefont {I.}~\bibnamefont {Kovalchuk}}, \bibinfo {author} {\bibfnamefont {B.}~\bibnamefont {Piening}}, \bibinfo {author} {\bibfnamefont {V.}~\bibnamefont {Labracherie}}, \bibinfo {author} {\bibfnamefont {A.}~\bibnamefont {Veyrat}}, \bibinfo {author} {\bibfnamefont {D.}~\bibnamefont {Wolf}}, \bibinfo {author} {\bibfnamefont {A.}~\bibnamefont {Lubk}}, \bibinfo {author} {\bibfnamefont {S.}~\bibnamefont {Subakti}}, \bibinfo {author} {\bibfnamefont {R.}~\bibnamefont {Giraud}}, \bibinfo {author} {\bibfnamefont {J.}~\bibnamefont {Dufouleur}}, \emph {et~al.},\ }\href@noop {} {\bibfield  {journal} {\bibinfo  {journal} {Physical Review Materials}\ }\textbf {\bibinfo {volume} {4}},\ \bibinfo {pages} {124202} (\bibinfo {year} {2020})}\BibitemShut {NoStop}%
\bibitem [{\citenamefont {Wu}\ \emph {et~al.}(2019)\citenamefont {Wu}, \citenamefont {Jo}, \citenamefont {Wang}, \citenamefont {Schmidt}, \citenamefont {Neilson}, \citenamefont {Schrunk}, \citenamefont {Swatek}, \citenamefont {Eaton}, \citenamefont {Bud'ko}, \citenamefont {Canfield},\ and\ \citenamefont {Kaminski}}]{PhysRevB.99.161113}%
  \BibitemOpen
  \bibfield  {author} {\bibinfo {author} {\bibfnamefont {Y.}~\bibnamefont {Wu}}, \bibinfo {author} {\bibfnamefont {N.~H.}\ \bibnamefont {Jo}}, \bibinfo {author} {\bibfnamefont {L.-L.}\ \bibnamefont {Wang}}, \bibinfo {author} {\bibfnamefont {C.~A.}\ \bibnamefont {Schmidt}}, \bibinfo {author} {\bibfnamefont {K.~M.}\ \bibnamefont {Neilson}}, \bibinfo {author} {\bibfnamefont {B.}~\bibnamefont {Schrunk}}, \bibinfo {author} {\bibfnamefont {P.}~\bibnamefont {Swatek}}, \bibinfo {author} {\bibfnamefont {A.}~\bibnamefont {Eaton}}, \bibinfo {author} {\bibfnamefont {S.~L.}\ \bibnamefont {Bud'ko}}, \bibinfo {author} {\bibfnamefont {P.~C.}\ \bibnamefont {Canfield}},\ and\ \bibinfo {author} {\bibfnamefont {A.}~\bibnamefont {Kaminski}},\ }\href {https://doi.org/10.1103/PhysRevB.99.161113} {\bibfield  {journal} {\bibinfo  {journal} {Phys. Rev. B}\ }\textbf {\bibinfo {volume} {99}},\ \bibinfo {pages} {161113} (\bibinfo {year} {2019})}\BibitemShut {NoStop}%
\bibitem [{\citenamefont {Kuibarov}\ \emph {et~al.}(2024)\citenamefont {Kuibarov}, \citenamefont {Suvorov}, \citenamefont {Vocaturo}, \citenamefont {Fedorov}, \citenamefont {Lou}, \citenamefont {Merkwitz}, \citenamefont {Voroshnin}, \citenamefont {Facio}, \citenamefont {Koepernik}, \citenamefont {Yaresko} \emph {et~al.}}]{kuibarov2024evidence}%
  \BibitemOpen
  \bibfield  {author} {\bibinfo {author} {\bibfnamefont {A.}~\bibnamefont {Kuibarov}}, \bibinfo {author} {\bibfnamefont {O.}~\bibnamefont {Suvorov}}, \bibinfo {author} {\bibfnamefont {R.}~\bibnamefont {Vocaturo}}, \bibinfo {author} {\bibfnamefont {A.}~\bibnamefont {Fedorov}}, \bibinfo {author} {\bibfnamefont {R.}~\bibnamefont {Lou}}, \bibinfo {author} {\bibfnamefont {L.}~\bibnamefont {Merkwitz}}, \bibinfo {author} {\bibfnamefont {V.}~\bibnamefont {Voroshnin}}, \bibinfo {author} {\bibfnamefont {J.~I.}\ \bibnamefont {Facio}}, \bibinfo {author} {\bibfnamefont {K.}~\bibnamefont {Koepernik}}, \bibinfo {author} {\bibfnamefont {A.}~\bibnamefont {Yaresko}}, \emph {et~al.},\ }\href@noop {} {\bibfield  {journal} {\bibinfo  {journal} {Nature}\ }\textbf {\bibinfo {volume} {626}},\ \bibinfo {pages} {294} (\bibinfo {year} {2024})}\BibitemShut {NoStop}%
\bibitem [{\citenamefont {Vocaturo}\ \emph {et~al.}(2024)\citenamefont {Vocaturo}, \citenamefont {Koepernik}, \citenamefont {Facio}, \citenamefont {Timm}, \citenamefont {Fulga}, \citenamefont {Janson},\ and\ \citenamefont {van~den Brink}}]{PhysRevB.110.054504}%
  \BibitemOpen
  \bibfield  {author} {\bibinfo {author} {\bibfnamefont {R.}~\bibnamefont {Vocaturo}}, \bibinfo {author} {\bibfnamefont {K.}~\bibnamefont {Koepernik}}, \bibinfo {author} {\bibfnamefont {J.~I.}\ \bibnamefont {Facio}}, \bibinfo {author} {\bibfnamefont {C.}~\bibnamefont {Timm}}, \bibinfo {author} {\bibfnamefont {I.~C.}\ \bibnamefont {Fulga}}, \bibinfo {author} {\bibfnamefont {O.}~\bibnamefont {Janson}},\ and\ \bibinfo {author} {\bibfnamefont {J.}~\bibnamefont {van~den Brink}},\ }\href {https://doi.org/10.1103/PhysRevB.110.054504} {\bibfield  {journal} {\bibinfo  {journal} {Phys. Rev. B}\ }\textbf {\bibinfo {volume} {110}},\ \bibinfo {pages} {054504} (\bibinfo {year} {2024})}\BibitemShut {NoStop}%
\bibitem [{\citenamefont {Veyrat}\ \emph {et~al.}(2023)\citenamefont {Veyrat}, \citenamefont {Labracherie}, \citenamefont {Bashlakov}, \citenamefont {Caglieris}, \citenamefont {Facio}, \citenamefont {Shipunov}, \citenamefont {Charvin}, \citenamefont {Acharya}, \citenamefont {Naidyuk}, \citenamefont {Giraud} \emph {et~al.}}]{veyrat2023berezinskii}%
  \BibitemOpen
  \bibfield  {author} {\bibinfo {author} {\bibfnamefont {A.}~\bibnamefont {Veyrat}}, \bibinfo {author} {\bibfnamefont {V.}~\bibnamefont {Labracherie}}, \bibinfo {author} {\bibfnamefont {D.~L.}\ \bibnamefont {Bashlakov}}, \bibinfo {author} {\bibfnamefont {F.}~\bibnamefont {Caglieris}}, \bibinfo {author} {\bibfnamefont {J.~I.}\ \bibnamefont {Facio}}, \bibinfo {author} {\bibfnamefont {G.}~\bibnamefont {Shipunov}}, \bibinfo {author} {\bibfnamefont {T.}~\bibnamefont {Charvin}}, \bibinfo {author} {\bibfnamefont {R.}~\bibnamefont {Acharya}}, \bibinfo {author} {\bibfnamefont {Y.}~\bibnamefont {Naidyuk}}, \bibinfo {author} {\bibfnamefont {R.}~\bibnamefont {Giraud}}, \emph {et~al.},\ }\href@noop {} {\bibfield  {journal} {\bibinfo  {journal} {Nano Letters}\ }\textbf {\bibinfo {volume} {23}},\ \bibinfo {pages} {1229} (\bibinfo {year} {2023})}\BibitemShut {NoStop}%
\bibitem [{\citenamefont {Schimmel}\ \emph {et~al.}(2024)\citenamefont {Schimmel}, \citenamefont {Fasano}, \citenamefont {Hoffmann}, \citenamefont {Besproswanny}, \citenamefont {Corredor~Bohorquez}, \citenamefont {Puig}, \citenamefont {Elshalem}, \citenamefont {Kalisky}, \citenamefont {Shipunov}, \citenamefont {Baumann} \emph {et~al.}}]{schimmel2024surface}%
  \BibitemOpen
  \bibfield  {author} {\bibinfo {author} {\bibfnamefont {S.}~\bibnamefont {Schimmel}}, \bibinfo {author} {\bibfnamefont {Y.}~\bibnamefont {Fasano}}, \bibinfo {author} {\bibfnamefont {S.}~\bibnamefont {Hoffmann}}, \bibinfo {author} {\bibfnamefont {J.}~\bibnamefont {Besproswanny}}, \bibinfo {author} {\bibfnamefont {L.~T.}\ \bibnamefont {Corredor~Bohorquez}}, \bibinfo {author} {\bibfnamefont {J.}~\bibnamefont {Puig}}, \bibinfo {author} {\bibfnamefont {B.-C.}\ \bibnamefont {Elshalem}}, \bibinfo {author} {\bibfnamefont {B.}~\bibnamefont {Kalisky}}, \bibinfo {author} {\bibfnamefont {G.}~\bibnamefont {Shipunov}}, \bibinfo {author} {\bibfnamefont {D.}~\bibnamefont {Baumann}}, \emph {et~al.},\ }\href@noop {} {\bibfield  {journal} {\bibinfo  {journal} {Nature Communications}\ }\textbf {\bibinfo {volume} {15}},\ \bibinfo {pages} {9895} (\bibinfo {year} {2024})}\BibitemShut {NoStop}%
\bibitem [{\citenamefont {Kaminski}\ \emph {et~al.}(2015)\citenamefont {Kaminski}, \citenamefont {Kondo}, \citenamefont {Takeuchi},\ and\ \citenamefont {Gu}}]{kaminski2015pairing}%
  \BibitemOpen
  \bibfield  {author} {\bibinfo {author} {\bibfnamefont {A.}~\bibnamefont {Kaminski}}, \bibinfo {author} {\bibfnamefont {T.}~\bibnamefont {Kondo}}, \bibinfo {author} {\bibfnamefont {T.}~\bibnamefont {Takeuchi}},\ and\ \bibinfo {author} {\bibfnamefont {G.}~\bibnamefont {Gu}},\ }\href@noop {} {\bibfield  {journal} {\bibinfo  {journal} {Philosophical Magazine}\ }\textbf {\bibinfo {volume} {95}},\ \bibinfo {pages} {453} (\bibinfo {year} {2015})}\BibitemShut {NoStop}%
\bibitem [{\citenamefont {Schrunk}\ \emph {et~al.}(2022)\citenamefont {Schrunk}, \citenamefont {Kushnirenko}, \citenamefont {Kuthanazhi}, \citenamefont {Ahn}, \citenamefont {Wang}, \citenamefont {O’Leary}, \citenamefont {Lee}, \citenamefont {Eaton}, \citenamefont {Fedorov}, \citenamefont {Lou} \emph {et~al.}}]{schrunk2022emergence}%
  \BibitemOpen
  \bibfield  {author} {\bibinfo {author} {\bibfnamefont {B.}~\bibnamefont {Schrunk}}, \bibinfo {author} {\bibfnamefont {Y.}~\bibnamefont {Kushnirenko}}, \bibinfo {author} {\bibfnamefont {B.}~\bibnamefont {Kuthanazhi}}, \bibinfo {author} {\bibfnamefont {J.}~\bibnamefont {Ahn}}, \bibinfo {author} {\bibfnamefont {L.-L.}\ \bibnamefont {Wang}}, \bibinfo {author} {\bibfnamefont {E.}~\bibnamefont {O’Leary}}, \bibinfo {author} {\bibfnamefont {K.}~\bibnamefont {Lee}}, \bibinfo {author} {\bibfnamefont {A.}~\bibnamefont {Eaton}}, \bibinfo {author} {\bibfnamefont {A.}~\bibnamefont {Fedorov}}, \bibinfo {author} {\bibfnamefont {R.}~\bibnamefont {Lou}}, \emph {et~al.},\ }\href@noop {} {\bibfield  {journal} {\bibinfo  {journal} {Nature}\ }\textbf {\bibinfo {volume} {603}},\ \bibinfo {pages} {610} (\bibinfo {year} {2022})}\BibitemShut {NoStop}%
\bibitem [{\citenamefont {Kushnirenko}\ \emph {et~al.}(2022)\citenamefont {Kushnirenko}, \citenamefont {Schrunk}, \citenamefont {Kuthanazhi}, \citenamefont {Wang}, \citenamefont {Ahn}, \citenamefont {O'Leary}, \citenamefont {Eaton}, \citenamefont {Bud'ko}, \citenamefont {Slager}, \citenamefont {Canfield} \emph {et~al.}}]{kushnirenko2022rare}%
  \BibitemOpen
  \bibfield  {author} {\bibinfo {author} {\bibfnamefont {Y.}~\bibnamefont {Kushnirenko}}, \bibinfo {author} {\bibfnamefont {B.}~\bibnamefont {Schrunk}}, \bibinfo {author} {\bibfnamefont {B.}~\bibnamefont {Kuthanazhi}}, \bibinfo {author} {\bibfnamefont {L.-L.}\ \bibnamefont {Wang}}, \bibinfo {author} {\bibfnamefont {J.}~\bibnamefont {Ahn}}, \bibinfo {author} {\bibfnamefont {E.}~\bibnamefont {O'Leary}}, \bibinfo {author} {\bibfnamefont {A.}~\bibnamefont {Eaton}}, \bibinfo {author} {\bibfnamefont {S.~L.}\ \bibnamefont {Bud'ko}}, \bibinfo {author} {\bibfnamefont {R.-J.}\ \bibnamefont {Slager}}, \bibinfo {author} {\bibfnamefont {P.}~\bibnamefont {Canfield}}, \emph {et~al.},\ }\href@noop {} {\bibfield  {journal} {\bibinfo  {journal} {Physical Review B}\ }\textbf {\bibinfo {volume} {106}},\ \bibinfo {pages} {115112} (\bibinfo {year} {2022})}\BibitemShut {NoStop}%
\bibitem [{\citenamefont {Kushnirenko}\ \emph {et~al.}(2024)\citenamefont {Kushnirenko}, \citenamefont {Schrunk}, \citenamefont {Kuthanazhi}, \citenamefont {Wang}, \citenamefont {Ahn}, \citenamefont {O'Leary}, \citenamefont {Eaton}, \citenamefont {Bud'ko}, \citenamefont {Slager}, \citenamefont {Canfield} \emph {et~al.}}]{kushnirenko2024unexpected}%
  \BibitemOpen
  \bibfield  {author} {\bibinfo {author} {\bibfnamefont {Y.}~\bibnamefont {Kushnirenko}}, \bibinfo {author} {\bibfnamefont {B.}~\bibnamefont {Schrunk}}, \bibinfo {author} {\bibfnamefont {B.}~\bibnamefont {Kuthanazhi}}, \bibinfo {author} {\bibfnamefont {L.-L.}\ \bibnamefont {Wang}}, \bibinfo {author} {\bibfnamefont {J.}~\bibnamefont {Ahn}}, \bibinfo {author} {\bibfnamefont {E.}~\bibnamefont {O'Leary}}, \bibinfo {author} {\bibfnamefont {A.}~\bibnamefont {Eaton}}, \bibinfo {author} {\bibfnamefont {S.~L.}\ \bibnamefont {Bud'ko}}, \bibinfo {author} {\bibfnamefont {R.-J.}\ \bibnamefont {Slager}}, \bibinfo {author} {\bibfnamefont {P.}~\bibnamefont {Canfield}}, \emph {et~al.},\ }\href@noop {} {\bibfield  {journal} {\bibinfo  {journal} {Nature Communications}\ } (\bibinfo {year} {2024})}\BibitemShut {NoStop}%
\bibitem [{\citenamefont {Kushnirenko}\ \emph {et~al.}(2023)\citenamefont {Kushnirenko}, \citenamefont {Kuthanazhi}, \citenamefont {Wang}, \citenamefont {Schrunk}, \citenamefont {O'Leary}, \citenamefont {Eaton}, \citenamefont {Canfield},\ and\ \citenamefont {Kaminski}}]{kushnirenko2023directional}%
  \BibitemOpen
  \bibfield  {author} {\bibinfo {author} {\bibfnamefont {Y.}~\bibnamefont {Kushnirenko}}, \bibinfo {author} {\bibfnamefont {B.}~\bibnamefont {Kuthanazhi}}, \bibinfo {author} {\bibfnamefont {L.-L.}\ \bibnamefont {Wang}}, \bibinfo {author} {\bibfnamefont {B.}~\bibnamefont {Schrunk}}, \bibinfo {author} {\bibfnamefont {E.}~\bibnamefont {O'Leary}}, \bibinfo {author} {\bibfnamefont {A.}~\bibnamefont {Eaton}}, \bibinfo {author} {\bibfnamefont {P.~C.}\ \bibnamefont {Canfield}},\ and\ \bibinfo {author} {\bibfnamefont {A.}~\bibnamefont {Kaminski}},\ }\href@noop {} {\bibfield  {journal} {\bibinfo  {journal} {Physical Review B}\ }\textbf {\bibinfo {volume} {108}},\ \bibinfo {pages} {115102} (\bibinfo {year} {2023})}\BibitemShut {NoStop}%
\bibitem [{\citenamefont {Lv}\ \emph {et~al.}(2015)\citenamefont {Lv}, \citenamefont {Weng}, \citenamefont {Fu}, \citenamefont {Wang}, \citenamefont {Miao}, \citenamefont {Ma}, \citenamefont {Richard}, \citenamefont {Huang}, \citenamefont {Zhao}, \citenamefont {Chen} \emph {et~al.}}]{lv2015experimental}%
  \BibitemOpen
  \bibfield  {author} {\bibinfo {author} {\bibfnamefont {B.}~\bibnamefont {Lv}}, \bibinfo {author} {\bibfnamefont {H.}~\bibnamefont {Weng}}, \bibinfo {author} {\bibfnamefont {B.}~\bibnamefont {Fu}}, \bibinfo {author} {\bibfnamefont {X.~P.}\ \bibnamefont {Wang}}, \bibinfo {author} {\bibfnamefont {H.}~\bibnamefont {Miao}}, \bibinfo {author} {\bibfnamefont {J.}~\bibnamefont {Ma}}, \bibinfo {author} {\bibfnamefont {P.}~\bibnamefont {Richard}}, \bibinfo {author} {\bibfnamefont {X.}~\bibnamefont {Huang}}, \bibinfo {author} {\bibfnamefont {L.}~\bibnamefont {Zhao}}, \bibinfo {author} {\bibfnamefont {G.}~\bibnamefont {Chen}}, \emph {et~al.},\ }\href@noop {} {\bibfield  {journal} {\bibinfo  {journal} {Physical Review X}\ }\textbf {\bibinfo {volume} {5}},\ \bibinfo {pages} {031013} (\bibinfo {year} {2015})}\BibitemShut {NoStop}%
\bibitem [{\citenamefont {Yang}\ \emph {et~al.}(2015)\citenamefont {Yang}, \citenamefont {Liu}, \citenamefont {Sun}, \citenamefont {Peng}, \citenamefont {Yang}, \citenamefont {Zhang}, \citenamefont {Zhou}, \citenamefont {Zhang}, \citenamefont {Guo}, \citenamefont {Rahn} \emph {et~al.}}]{yang2015weyl}%
  \BibitemOpen
  \bibfield  {author} {\bibinfo {author} {\bibfnamefont {L.}~\bibnamefont {Yang}}, \bibinfo {author} {\bibfnamefont {Z.}~\bibnamefont {Liu}}, \bibinfo {author} {\bibfnamefont {Y.}~\bibnamefont {Sun}}, \bibinfo {author} {\bibfnamefont {H.}~\bibnamefont {Peng}}, \bibinfo {author} {\bibfnamefont {H.}~\bibnamefont {Yang}}, \bibinfo {author} {\bibfnamefont {T.}~\bibnamefont {Zhang}}, \bibinfo {author} {\bibfnamefont {B.}~\bibnamefont {Zhou}}, \bibinfo {author} {\bibfnamefont {Y.}~\bibnamefont {Zhang}}, \bibinfo {author} {\bibfnamefont {Y.}~\bibnamefont {Guo}}, \bibinfo {author} {\bibfnamefont {M.}~\bibnamefont {Rahn}}, \emph {et~al.},\ }\href@noop {} {\bibfield  {journal} {\bibinfo  {journal} {Nature physics}\ }\textbf {\bibinfo {volume} {11}},\ \bibinfo {pages} {728} (\bibinfo {year} {2015})}\BibitemShut {NoStop}%
\bibitem [{\citenamefont {Xu}\ \emph {et~al.}(2016)\citenamefont {Xu}, \citenamefont {Weng}, \citenamefont {Lv}, \citenamefont {Matt}, \citenamefont {Park}, \citenamefont {Bisti}, \citenamefont {Strocov}, \citenamefont {Gawryluk}, \citenamefont {Pomjakushina}, \citenamefont {Conder} \emph {et~al.}}]{xu2016observation}%
  \BibitemOpen
  \bibfield  {author} {\bibinfo {author} {\bibfnamefont {N.}~\bibnamefont {Xu}}, \bibinfo {author} {\bibfnamefont {H.}~\bibnamefont {Weng}}, \bibinfo {author} {\bibfnamefont {B.}~\bibnamefont {Lv}}, \bibinfo {author} {\bibfnamefont {C.~E.}\ \bibnamefont {Matt}}, \bibinfo {author} {\bibfnamefont {J.}~\bibnamefont {Park}}, \bibinfo {author} {\bibfnamefont {F.}~\bibnamefont {Bisti}}, \bibinfo {author} {\bibfnamefont {V.~N.}\ \bibnamefont {Strocov}}, \bibinfo {author} {\bibfnamefont {D.}~\bibnamefont {Gawryluk}}, \bibinfo {author} {\bibfnamefont {E.}~\bibnamefont {Pomjakushina}}, \bibinfo {author} {\bibfnamefont {K.}~\bibnamefont {Conder}}, \emph {et~al.},\ }\href@noop {} {\bibfield  {journal} {\bibinfo  {journal} {Nature communications}\ }\textbf {\bibinfo {volume} {7}},\ \bibinfo {pages} {11006} (\bibinfo {year} {2016})}\BibitemShut {NoStop}%
\bibitem [{\citenamefont {Xu}\ \emph {et~al.}(2015)\citenamefont {Xu}, \citenamefont {Alidoust}, \citenamefont {Belopolski}, \citenamefont {Yuan}, \citenamefont {Bian}, \citenamefont {Chang}, \citenamefont {Zheng}, \citenamefont {Strocov}, \citenamefont {Sanchez}, \citenamefont {Chang} \emph {et~al.}}]{xu2015discovery}%
  \BibitemOpen
  \bibfield  {author} {\bibinfo {author} {\bibfnamefont {S.-Y.}\ \bibnamefont {Xu}}, \bibinfo {author} {\bibfnamefont {N.}~\bibnamefont {Alidoust}}, \bibinfo {author} {\bibfnamefont {I.}~\bibnamefont {Belopolski}}, \bibinfo {author} {\bibfnamefont {Z.}~\bibnamefont {Yuan}}, \bibinfo {author} {\bibfnamefont {G.}~\bibnamefont {Bian}}, \bibinfo {author} {\bibfnamefont {T.-R.}\ \bibnamefont {Chang}}, \bibinfo {author} {\bibfnamefont {H.}~\bibnamefont {Zheng}}, \bibinfo {author} {\bibfnamefont {V.~N.}\ \bibnamefont {Strocov}}, \bibinfo {author} {\bibfnamefont {D.~S.}\ \bibnamefont {Sanchez}}, \bibinfo {author} {\bibfnamefont {G.}~\bibnamefont {Chang}}, \emph {et~al.},\ }\href@noop {} {\bibfield  {journal} {\bibinfo  {journal} {Nature Physics}\ }\textbf {\bibinfo {volume} {11}},\ \bibinfo {pages} {748} (\bibinfo {year} {2015})}\BibitemShut {NoStop}%
\bibitem [{\citenamefont {Souma}\ \emph {et~al.}(2016)\citenamefont {Souma}, \citenamefont {Wang}, \citenamefont {Kotaka}, \citenamefont {Sato}, \citenamefont {Nakayama}, \citenamefont {Tanaka}, \citenamefont {Kimizuka}, \citenamefont {Takahashi}, \citenamefont {Yamauchi}, \citenamefont {Oguchi} \emph {et~al.}}]{souma2016direct}%
  \BibitemOpen
  \bibfield  {author} {\bibinfo {author} {\bibfnamefont {S.}~\bibnamefont {Souma}}, \bibinfo {author} {\bibfnamefont {Z.}~\bibnamefont {Wang}}, \bibinfo {author} {\bibfnamefont {H.}~\bibnamefont {Kotaka}}, \bibinfo {author} {\bibfnamefont {T.}~\bibnamefont {Sato}}, \bibinfo {author} {\bibfnamefont {K.}~\bibnamefont {Nakayama}}, \bibinfo {author} {\bibfnamefont {Y.}~\bibnamefont {Tanaka}}, \bibinfo {author} {\bibfnamefont {H.}~\bibnamefont {Kimizuka}}, \bibinfo {author} {\bibfnamefont {T.}~\bibnamefont {Takahashi}}, \bibinfo {author} {\bibfnamefont {K.}~\bibnamefont {Yamauchi}}, \bibinfo {author} {\bibfnamefont {T.}~\bibnamefont {Oguchi}}, \emph {et~al.},\ }\href@noop {} {\bibfield  {journal} {\bibinfo  {journal} {Physical Review B}\ }\textbf {\bibinfo {volume} {93}},\ \bibinfo {pages} {161112} (\bibinfo {year} {2016})}\BibitemShut {NoStop}%
\bibitem [{\citenamefont {Lee}\ \emph {et~al.}(2021)\citenamefont {Lee}, \citenamefont {Lange}, \citenamefont {Wang}, \citenamefont {Kuthanazhi}, \citenamefont {Trevisan}, \citenamefont {Jo}, \citenamefont {Schrunk}, \citenamefont {Orth}, \citenamefont {Slager}, \citenamefont {Canfield} \emph {et~al.}}]{lee2021discovery}%
  \BibitemOpen
  \bibfield  {author} {\bibinfo {author} {\bibfnamefont {K.}~\bibnamefont {Lee}}, \bibinfo {author} {\bibfnamefont {G.~F.}\ \bibnamefont {Lange}}, \bibinfo {author} {\bibfnamefont {L.-L.}\ \bibnamefont {Wang}}, \bibinfo {author} {\bibfnamefont {B.}~\bibnamefont {Kuthanazhi}}, \bibinfo {author} {\bibfnamefont {T.~V.}\ \bibnamefont {Trevisan}}, \bibinfo {author} {\bibfnamefont {N.~H.}\ \bibnamefont {Jo}}, \bibinfo {author} {\bibfnamefont {B.}~\bibnamefont {Schrunk}}, \bibinfo {author} {\bibfnamefont {P.~P.}\ \bibnamefont {Orth}}, \bibinfo {author} {\bibfnamefont {R.-J.}\ \bibnamefont {Slager}}, \bibinfo {author} {\bibfnamefont {P.~C.}\ \bibnamefont {Canfield}}, \emph {et~al.},\ }\href@noop {} {\bibfield  {journal} {\bibinfo  {journal} {Nature communications}\ }\textbf {\bibinfo {volume} {12}},\ \bibinfo {pages} {1855} (\bibinfo {year} {2021})}\BibitemShut {NoStop}%
\bibitem [{\citenamefont {Canfield}\ \emph {et~al.}(2016)\citenamefont {Canfield}, \citenamefont {Kong}, \citenamefont {Kaluarachchi},\ and\ \citenamefont {Jo}}]{canfield2016use}%
  \BibitemOpen
  \bibfield  {author} {\bibinfo {author} {\bibfnamefont {P.~C.}\ \bibnamefont {Canfield}}, \bibinfo {author} {\bibfnamefont {T.}~\bibnamefont {Kong}}, \bibinfo {author} {\bibfnamefont {U.~S.}\ \bibnamefont {Kaluarachchi}},\ and\ \bibinfo {author} {\bibfnamefont {N.~H.}\ \bibnamefont {Jo}},\ }\href@noop {} {\bibfield  {journal} {\bibinfo  {journal} {Philosophical magazine}\ }\textbf {\bibinfo {volume} {96}},\ \bibinfo {pages} {84} (\bibinfo {year} {2016})}\BibitemShut {NoStop}%
\bibitem [{\citenamefont {Canfield}(2015)}]{crucible}%
  \BibitemOpen
  \bibfield  {author} {\bibinfo {author} {\bibfnamefont {P.~C.}\ \bibnamefont {Canfield}},\ }\href {https://www.lspceramics.com/canfield-crucible-sets-2/} {\bibinfo {title} {Canfield crucible set, https://www.lspceramics.com/canfield-crucible-sets-2/}} (\bibinfo {year} {2015})\BibitemShut {NoStop}%
\bibitem [{\citenamefont {Canfield}(2019)}]{canfield2019new}%
  \BibitemOpen
  \bibfield  {author} {\bibinfo {author} {\bibfnamefont {P.~C.}\ \bibnamefont {Canfield}},\ }\href@noop {} {\bibfield  {journal} {\bibinfo  {journal} {Reports on Progress in Physics}\ }\textbf {\bibinfo {volume} {83}},\ \bibinfo {pages} {016501} (\bibinfo {year} {2019})}\BibitemShut {NoStop}%
\bibitem [{\citenamefont {Jiang}\ \emph {et~al.}(2014)\citenamefont {Jiang}, \citenamefont {Mou}, \citenamefont {Wu}, \citenamefont {Huang}, \citenamefont {McMillen}, \citenamefont {Kolis}, \citenamefont {Giesber~III}, \citenamefont {Egan},\ and\ \citenamefont {Kaminski}}]{jiang2014tunable}%
  \BibitemOpen
  \bibfield  {author} {\bibinfo {author} {\bibfnamefont {R.}~\bibnamefont {Jiang}}, \bibinfo {author} {\bibfnamefont {D.}~\bibnamefont {Mou}}, \bibinfo {author} {\bibfnamefont {Y.}~\bibnamefont {Wu}}, \bibinfo {author} {\bibfnamefont {L.}~\bibnamefont {Huang}}, \bibinfo {author} {\bibfnamefont {C.~D.}\ \bibnamefont {McMillen}}, \bibinfo {author} {\bibfnamefont {J.}~\bibnamefont {Kolis}}, \bibinfo {author} {\bibfnamefont {H.~G.}\ \bibnamefont {Giesber~III}}, \bibinfo {author} {\bibfnamefont {J.~J.}\ \bibnamefont {Egan}},\ and\ \bibinfo {author} {\bibfnamefont {A.}~\bibnamefont {Kaminski}},\ }\href@noop {} {\bibfield  {journal} {\bibinfo  {journal} {Review of Scientific Instruments}\ }\textbf {\bibinfo {volume} {85}},\ \bibinfo {pages} {033902} (\bibinfo {year} {2014})}\BibitemShut {NoStop}%
\bibitem [{\citenamefont {Hohenberg}\ and\ \citenamefont {Kohn}(1964)}]{hohenberg1964inhomogeneous_DFT1}%
  \BibitemOpen
  \bibfield  {author} {\bibinfo {author} {\bibfnamefont {P.}~\bibnamefont {Hohenberg}}\ and\ \bibinfo {author} {\bibfnamefont {W.}~\bibnamefont {Kohn}},\ }\href@noop {} {\bibfield  {journal} {\bibinfo  {journal} {Physical review}\ }\textbf {\bibinfo {volume} {136}},\ \bibinfo {pages} {B864} (\bibinfo {year} {1964})}\BibitemShut {NoStop}%
\bibitem [{\citenamefont {Kohn}\ and\ \citenamefont {Sham}(1965)}]{kohn1965self_DFT2}%
  \BibitemOpen
  \bibfield  {author} {\bibinfo {author} {\bibfnamefont {W.}~\bibnamefont {Kohn}}\ and\ \bibinfo {author} {\bibfnamefont {L.~J.}\ \bibnamefont {Sham}},\ }\href@noop {} {\bibfield  {journal} {\bibinfo  {journal} {Physical Review}\ }\textbf {\bibinfo {volume} {140}},\ \bibinfo {pages} {A1133} (\bibinfo {year} {1965})}\BibitemShut {NoStop}%
\bibitem [{\citenamefont {Perdew}\ \emph {et~al.}(1996)\citenamefont {Perdew}, \citenamefont {Burke},\ and\ \citenamefont {Ernzerhof}}]{perdew1996generalized_DFT3}%
  \BibitemOpen
  \bibfield  {author} {\bibinfo {author} {\bibfnamefont {J.~P.}\ \bibnamefont {Perdew}}, \bibinfo {author} {\bibfnamefont {K.}~\bibnamefont {Burke}},\ and\ \bibinfo {author} {\bibfnamefont {M.}~\bibnamefont {Ernzerhof}},\ }\href@noop {} {\bibfield  {journal} {\bibinfo  {journal} {Physical Review Letters}\ }\textbf {\bibinfo {volume} {77}},\ \bibinfo {pages} {3865} (\bibinfo {year} {1996})}\BibitemShut {NoStop}%
\bibitem [{\citenamefont {Bl{\"o}chl}(1994)}]{blochl1994projector_DFT4}%
  \BibitemOpen
  \bibfield  {author} {\bibinfo {author} {\bibfnamefont {P.~E.}\ \bibnamefont {Bl{\"o}chl}},\ }\href@noop {} {\bibfield  {journal} {\bibinfo  {journal} {Physical Review B}\ }\textbf {\bibinfo {volume} {50}},\ \bibinfo {pages} {17953} (\bibinfo {year} {1994})}\BibitemShut {NoStop}%
\bibitem [{\citenamefont {Kresse}\ and\ \citenamefont {Furthm{\"u}ller}(1996{\natexlab{a}})}]{kresse1996efficient_DFT5}%
  \BibitemOpen
  \bibfield  {author} {\bibinfo {author} {\bibfnamefont {G.}~\bibnamefont {Kresse}}\ and\ \bibinfo {author} {\bibfnamefont {J.}~\bibnamefont {Furthm{\"u}ller}},\ }\href@noop {} {\bibfield  {journal} {\bibinfo  {journal} {Physical Review B}\ }\textbf {\bibinfo {volume} {54}},\ \bibinfo {pages} {11169} (\bibinfo {year} {1996}{\natexlab{a}})}\BibitemShut {NoStop}%
\bibitem [{\citenamefont {Kresse}\ and\ \citenamefont {Furthm{\"u}ller}(1996{\natexlab{b}})}]{kresse1996efficiency_DFT6}%
  \BibitemOpen
  \bibfield  {author} {\bibinfo {author} {\bibfnamefont {G.}~\bibnamefont {Kresse}}\ and\ \bibinfo {author} {\bibfnamefont {J.}~\bibnamefont {Furthm{\"u}ller}},\ }\href@noop {} {\bibfield  {journal} {\bibinfo  {journal} {Computational Materials Science}\ }\textbf {\bibinfo {volume} {6}},\ \bibinfo {pages} {15} (\bibinfo {year} {1996}{\natexlab{b}})}\BibitemShut {NoStop}%
\bibitem [{\citenamefont {Kaiser}\ \emph {et~al.}(2014)\citenamefont {Kaiser}, \citenamefont {Baranov},\ and\ \citenamefont {Ruck}}]{kaiser2014bi2pt}%
  \BibitemOpen
  \bibfield  {author} {\bibinfo {author} {\bibfnamefont {M.}~\bibnamefont {Kaiser}}, \bibinfo {author} {\bibfnamefont {A.~I.}\ \bibnamefont {Baranov}},\ and\ \bibinfo {author} {\bibfnamefont {M.}~\bibnamefont {Ruck}},\ }\href@noop {} {\bibfield  {journal} {\bibinfo  {journal} {Zeitschrift f{\"u}r anorganische und allgemeine Chemie}\ }\textbf {\bibinfo {volume} {640}},\ \bibinfo {pages} {2742} (\bibinfo {year} {2014})}\BibitemShut {NoStop}%
\bibitem [{\citenamefont {Monkhorst}\ and\ \citenamefont {Pack}(1976)}]{monkhorst1976special_DFT8}%
  \BibitemOpen
  \bibfield  {author} {\bibinfo {author} {\bibfnamefont {H.~J.}\ \bibnamefont {Monkhorst}}\ and\ \bibinfo {author} {\bibfnamefont {J.~D.}\ \bibnamefont {Pack}},\ }\href@noop {} {\bibfield  {journal} {\bibinfo  {journal} {Physical Review B}\ }\textbf {\bibinfo {volume} {13}},\ \bibinfo {pages} {5188} (\bibinfo {year} {1976})}\BibitemShut {NoStop}%
\bibitem [{\citenamefont {Marzari}\ and\ \citenamefont {Vanderbilt}(1997)}]{marzari1997maximally_DFT9}%
  \BibitemOpen
  \bibfield  {author} {\bibinfo {author} {\bibfnamefont {N.}~\bibnamefont {Marzari}}\ and\ \bibinfo {author} {\bibfnamefont {D.}~\bibnamefont {Vanderbilt}},\ }\href@noop {} {\bibfield  {journal} {\bibinfo  {journal} {Physical Review B}\ }\textbf {\bibinfo {volume} {56}},\ \bibinfo {pages} {12847} (\bibinfo {year} {1997})}\BibitemShut {NoStop}%
\bibitem [{\citenamefont {Souza}\ \emph {et~al.}(2001)\citenamefont {Souza}, \citenamefont {Marzari},\ and\ \citenamefont {Vanderbilt}}]{souza2001maximally_DFT10}%
  \BibitemOpen
  \bibfield  {author} {\bibinfo {author} {\bibfnamefont {I.}~\bibnamefont {Souza}}, \bibinfo {author} {\bibfnamefont {N.}~\bibnamefont {Marzari}},\ and\ \bibinfo {author} {\bibfnamefont {D.}~\bibnamefont {Vanderbilt}},\ }\href@noop {} {\bibfield  {journal} {\bibinfo  {journal} {Physical Review B}\ }\textbf {\bibinfo {volume} {65}},\ \bibinfo {pages} {035109} (\bibinfo {year} {2001})}\BibitemShut {NoStop}%
\bibitem [{\citenamefont {Marzari}\ \emph {et~al.}(2012)\citenamefont {Marzari}, \citenamefont {Mostofi}, \citenamefont {Yates}, \citenamefont {Souza},\ and\ \citenamefont {Vanderbilt}}]{marzari2012maximally_DFT11}%
  \BibitemOpen
  \bibfield  {author} {\bibinfo {author} {\bibfnamefont {N.}~\bibnamefont {Marzari}}, \bibinfo {author} {\bibfnamefont {A.~A.}\ \bibnamefont {Mostofi}}, \bibinfo {author} {\bibfnamefont {J.~R.}\ \bibnamefont {Yates}}, \bibinfo {author} {\bibfnamefont {I.}~\bibnamefont {Souza}},\ and\ \bibinfo {author} {\bibfnamefont {D.}~\bibnamefont {Vanderbilt}},\ }\href@noop {} {\bibfield  {journal} {\bibinfo  {journal} {Reviews of Modern Physics}\ }\textbf {\bibinfo {volume} {84}},\ \bibinfo {pages} {1419} (\bibinfo {year} {2012})}\BibitemShut {NoStop}%
\bibitem [{\citenamefont {Sancho}\ \emph {et~al.}(1984)\citenamefont {Sancho}, \citenamefont {Sancho},\ and\ \citenamefont {Rubio}}]{sancho1984quick_DFT14}%
  \BibitemOpen
  \bibfield  {author} {\bibinfo {author} {\bibfnamefont {M.~L.}\ \bibnamefont {Sancho}}, \bibinfo {author} {\bibfnamefont {J.~L.}\ \bibnamefont {Sancho}},\ and\ \bibinfo {author} {\bibfnamefont {J.}~\bibnamefont {Rubio}},\ }\href@noop {} {\bibfield  {journal} {\bibinfo  {journal} {Journal of Physics F: Metal Physics}\ }\textbf {\bibinfo {volume} {14}},\ \bibinfo {pages} {1205} (\bibinfo {year} {1984})}\BibitemShut {NoStop}%
\bibitem [{\citenamefont {Sancho}\ \emph {et~al.}(1985)\citenamefont {Sancho}, \citenamefont {Sancho}, \citenamefont {Sancho},\ and\ \citenamefont {Rubio}}]{sancho1985highly_DFT15}%
  \BibitemOpen
  \bibfield  {author} {\bibinfo {author} {\bibfnamefont {M.~L.}\ \bibnamefont {Sancho}}, \bibinfo {author} {\bibfnamefont {J.~L.}\ \bibnamefont {Sancho}}, \bibinfo {author} {\bibfnamefont {J.~L.}\ \bibnamefont {Sancho}},\ and\ \bibinfo {author} {\bibfnamefont {J.}~\bibnamefont {Rubio}},\ }\href@noop {} {\bibfield  {journal} {\bibinfo  {journal} {Journal of Physics F: Metal Physics}\ }\textbf {\bibinfo {volume} {15}},\ \bibinfo {pages} {851} (\bibinfo {year} {1985})}\BibitemShut {NoStop}%
\bibitem [{\citenamefont {Wu}\ \emph {et~al.}(2018)\citenamefont {Wu}, \citenamefont {Zhang}, \citenamefont {Song}, \citenamefont {Troyer},\ and\ \citenamefont {Soluyanov}}]{wu2018wanniertools_DFT16}%
  \BibitemOpen
  \bibfield  {author} {\bibinfo {author} {\bibfnamefont {Q.}~\bibnamefont {Wu}}, \bibinfo {author} {\bibfnamefont {S.}~\bibnamefont {Zhang}}, \bibinfo {author} {\bibfnamefont {H.-F.}\ \bibnamefont {Song}}, \bibinfo {author} {\bibfnamefont {M.}~\bibnamefont {Troyer}},\ and\ \bibinfo {author} {\bibfnamefont {A.~A.}\ \bibnamefont {Soluyanov}},\ }\href@noop {} {\bibfield  {journal} {\bibinfo  {journal} {Computer Physics Communications}\ }\textbf {\bibinfo {volume} {224}},\ \bibinfo {pages} {405} (\bibinfo {year} {2018})}\BibitemShut {NoStop}%
\bibitem [{\citenamefont {Graf}\ \emph {et~al.}(2010)\citenamefont {Graf}, \citenamefont {Hellmann}, \citenamefont {Jozwiak}, \citenamefont {Smallwood}, \citenamefont {Hussain}, \citenamefont {Kaindl}, \citenamefont {Kipp}, \citenamefont {Rossnagel},\ and\ \citenamefont {Lanzara}}]{graf2010vacuum}%
  \BibitemOpen
  \bibfield  {author} {\bibinfo {author} {\bibfnamefont {J.}~\bibnamefont {Graf}}, \bibinfo {author} {\bibfnamefont {S.}~\bibnamefont {Hellmann}}, \bibinfo {author} {\bibfnamefont {C.}~\bibnamefont {Jozwiak}}, \bibinfo {author} {\bibfnamefont {C.}~\bibnamefont {Smallwood}}, \bibinfo {author} {\bibfnamefont {Z.}~\bibnamefont {Hussain}}, \bibinfo {author} {\bibfnamefont {R.}~\bibnamefont {Kaindl}}, \bibinfo {author} {\bibfnamefont {L.}~\bibnamefont {Kipp}}, \bibinfo {author} {\bibfnamefont {K.}~\bibnamefont {Rossnagel}},\ and\ \bibinfo {author} {\bibfnamefont {A.}~\bibnamefont {Lanzara}},\ }\href@noop {} {\bibfield  {journal} {\bibinfo  {journal} {Journal of applied physics}\ }\textbf {\bibinfo {volume} {107}} (\bibinfo {year} {2010})}\BibitemShut {NoStop}%
\end{thebibliography}%

\end{document}